# Pressure-Tuned Intralayer Exchange in Superlattice-Like MnBi$_2$Te$_4$/(Bi$_2$Te$_3$)$_n$ Topological Insulators

Jifeng Shao,[†,∥] Yuntian Liu,[†,∥] Meng Zeng,[†] Jingyuan Li,[†] Xuefeng Wu,[†] Xiao-Ming Ma,[†] Feng Jin,[‡] Ruie Lu,[†] Yichen Sun,[†] Mingqiang Gu,[†] Kedong Wang,[†] Wenbin Wu,[‡] Liusuo Wu,[†] Chang Liu,[†] Qihang Liu,[*,†,§,#] and Yue Zhao[*,†]

[†]Shenzhen Institute for Quantum Science and Engineering and Department of Physics, Southern University of Science and Technology, Shenzhen 518055, China

[‡]Hefei National Laboratory for Physical Sciences at Microscale, University of Science and Technology of China, Hefei 230026, China

[§]Guangdong Provincial Key Laboratory for Computational Science and Material Design, Southern University of Science and Technology, Shenzhen 518055, China

[#]Shenzhen Key Laboratory of for Advanced Quantum Functional Materials and Devices, Southern University of Science and Technology, Shenzhen 518055, China

**ABSTRACT:**

The magnetic structures of MnBi$_2$Te$_4$(Bi$_2$Te$_3$)$_n$ can be manipulated by tuning the interlayer coupling via the number of Bi$_2$Te$_3$ spacer layers $n$, while the intralayer ferromagnetic (FM) exchange coupling is considered too robust to control. By applying hydrostatic pressure up to 3.5 GPa, we discover opposite responses of magnetic properties for $n$ = 1 and 2. MnBi$_4$Te$_7$ stays at A-type antiferromagnetic (AFM) phase with a decreasing Néel temperature and an increasing saturation field. In sharp contrast, MnBi$_6$Te$_{10}$ experiences a phase transition from A-type AFM to a quasi-two-dimensional FM state with a suppressed saturation field under pressure. First-principles calculations reveal the essential role of intralayer exchange coupling from lattice compression in determining these magnetic properties. Such magnetic phase transition is also observed in 20% Sb-doped MnBi$_6$Te$_{10}$ due to the in-plane lattice compression.

[*]To whom correspondence should be addressed: zhaoy@sustech.edu.cn, liuqh@sustech.edu.cn.

In magnetic topological insulators (MTI), the interplay between magnetic order in real space and the topological electronic structure in momentum space gives rise to many novel topological matters and emergent quantum phenomena, such as Weyl fermions, quantum anomalous Hall effect (QAHE), axion insulator phase and chiral Majorana modes[1-6]. A prototypical example is the layered intrinsic MTI $MnBi_2Te_4$ with the local moments of Mn atoms ferromagnetic (FM) aligned within one layer while adopting an A-type antiferromagnetic (AFM) order along the stacking direction[7-12]. By manipulating the magnetic order in few-layer $MnBi_2Te_4$ using film thickness and magnetic field, various topological phases have been experimentally realized, including zero-field QAHE and tunable axion insulator and other high order Chern insulator phases[13-19].

In the family of $MnBi_2Te_4(Bi_2Te_3)_n$, the interplay of magnetism and topology can be further enriched by inserting n layers of non-magnetic topological insulator (TI) $Bi_2Te_3$ into the van der Waals layered MTI $MnBi_2Te_4$[20-32]. Such heterostructure engineering of the non-magnetic TI and MTI building blocks not only reveals termination-dependent surface states and hybridization between different building blocks[25], but also effectively tunes the interlayer exchange coupling (IEC) between the neighboring magnetic layers, leading to new topological phases associated with different magnetic phases[27, 30, 31]. For n = 1, the transport and magnetism study of $MnBi_4Te_7$ shows A-type AFM state right below Néel temperature ($T_N$) and a competing magnetic order of FM state at even lower temperature (T ~ 1K)[22]. As n goes to 2, with a further weakened interlayer coupling between the magnetic layers, $MnBi_6Te_{10}$ keeps a relatively weak A-type AFM ground state behavior, with enhanced hysteresis loops in the magnetization curves below 6K [29]. Such magnetic behaviors could be associated with the competition between the weak interlayer exchange coupling and other magnetic interactions[22, 32]. Despite the complicated magnetism, a quasi-two-dimensional (2D) ferromagnet is proposed to dominate the magnetic properties at large n (n≥3) with a vanishing interlayer coupling[30, 31]. On the other hand, the intralayer exchange coupling is typically considered as robust FM, and thus difficult to manipulate by experimentally accessible "knobs".

As a clean, non-intrusive, reversible, and continuous structure tuning technique, hydrostatic pressure is usually utilized to modify the interlayer coupling strength by adjusting the interlayer separation in van der Waals structures[33-36]. In this work, by applying hydrostatic pressure up to 3.5 GPa, we found in $MnBi_2Te_4(Bi_2Te_3)_n$ (n = 1 and 2) distinct evolution of magnetic properties originated from the manipulation of intralayer exchange coupling. In contrast to the increasing saturation field in MnBi$_4$Te$_7$ (n = 1), the saturation field of MnBi$_6$Te$_{10}$ (n = 2) decreases by half to about 0.1 T at high pressure. In particular, an FM-like hysteresis of anomalous Hall resistivity and magnetic susceptibility emerges at 1.98 GPa, where the kinks or plateaus associated with AFM state between the polarized FM states vanish completely, suggesting the formation of FM domains. Our first-principles calculation shows that although the interlayer exchange coupling increases with pressure for both n = 1 and 2, hydrostatic pressure reduces the intralayer FM exchange coupling by evoking the competition between the AFM-preferred direct exchange and FM-preferred super-exchange coupling within the MnBi$_2$Te$_4$ layer. Considering the nearly vanishing interlayer coupling in MnBi$_6$Te$_{10}$ (about 0.01 meV, one order less compared with MnBi$_4$Te$_7$), the weakening of the intralayer FM exchange coupling will enhance the effect of fluctuation, including magnetic domains, thermal fluctuation, and other magnetic perturbations. The reduced intralayer magnetization can in turn decrease the interlayer exchange coupling strength (IEC ∝ $m_s$) in pressurized MnBi$_6$Te$_{10}$, and effectively decouple the adjacent magnetic layers, leading to a magnetic phase transition from weak A-type AFM to quasi-2D-FM state. A similar FM state is also observed in Sb-doped MnBi$_6$Te$_{10}$ with comparable in-plane lattice compression, providing additional evidence that reduced intralayer FM coupling can decouple the magnetic layers in weakly interlayer coupled MTI. Our results first reveal the delicate role of intralayer exchange coupling in the complex magnetic properties in MTI/TI heterostructures with a variable number of non-magnetic layers.

The single crystals were grown by flux method and first characterized by single crystal XRD on both the top and bottom surfaces (Figure S1). Subsequent screening using magnetic and magneto-transport measurements at ambient pressure was operated before applying high pressure (Figure S2 and S3). The measurement details under high pressure are described in the Supporting Information.

Figure 1a and 1b present the horizontal and Hall resistivity of MnBi$_4$Te$_7$ at various pressure at $T$ = 2 K. As pressure increases gradually to 3.39 GPa, the saturation field $B_s$ increases from 2.07 to 3.08 KOe and the hysteresis loops are strongly inhibited, indicating enhanced AFM interlayer exchange coupling. On the other hand, with increasing pressure, we observe a monotonic decrease of Néel temperature from 12.7 K to 9 K, directly contrasting to what could be expected at an enhanced AFM interlayer exchange coupling.

Figure 1e-g show the pressure-tuned transport behavior of MnBi$_6$Te$_{10}$. At ambient pressure, the weakened kinks around zero moment in the hysteresis loop mark the reduced interlayer AFM exchange coupling in MnBi$_6$Te$_{10}$ as n increases. Similar to MnBi$_4$Te$_7$, the magnetic phase transition temperature of MnBi$_6$Te$_{10}$ also decreases with increasing pressure. However, as pressure increases, the saturation field of MnBi$_6$Te$_{10}$ gradually decreases from 0.2 T to 0.08 T (Figure 1g). The responses to pressure of both the saturation field and magnetic ordering temperature of MnBi$_6$Te$_{10}$ are reversible, as shown in Figure S4d. The changes are inconsistent with the expected enhanced AFM interlayer coupling with compression along the *c*-axis. Note that the magnetic ordering temperature of MnBi$_4$Te$_7$ (~ 13K)[22, 24], MnBi$_6$Te$_{10}$ (~ 11K)[27-29], and MnBi$_8$Te$_{13}$ (~ 10.5K)[30, 31] are close to that of monolayer MnBi$_2$Te$_4$ (~ 12K)[14], the decreased magnetic ordering temperature could be related to the lattice compression within the magnetic septuple layer, since hydrostatic pressure simultaneously compresses the in-plane and out-of-plane lattice parameters.

Moreover, a relatively pure FM state emerges at pressure above 1.39 GPa, evidenced by a butterfly-shaped magnetoresistance and disappeared kinks associated with the weak AFM states in anomalous Hall resistance curves. Since no structural phase transitions are expected under pressure below 6 GPa in Bi$_2$Te$_3$, MnBi$_2$Te$_4$, and MnBi$_4$Te$_7$ single crystals[37-39], it is quite counterintuitive that an FM state emerges as the pressure-induced compression along *c*-axis shall strengthen the AFM interlayer exchange coupling.

To confirm the magnetic phase transition in MnBi$_6$Te$_{10}$, we applied the magnetic susceptibility and magnetization measurements under high pressure using a pre-calibrated Hall sensor. Measurement details are described in the Supporting Information. Figure 2 shows the temperature dependence of the magnetic susceptibility and the magnetization curves of MnBi$_6$Te$_{10}$ under different pressure with *H // c*. The *M-T* curve at ambient pressure measured by our method shows bifurcation of the zero-field cooling (ZFC) and field cooling (FC) at $T_N$, which matches well with the *M-T* curves directly measured by VSM, see Figure S8b. As pressure increases, the ZFC curve changes from a sharp cusp-like shape to a flat dome with increasing FC susceptibility below the magnetic phase transition temperature. The number of peaks on *dM/dH - H* curve reduces to 2 at 1.98 GPa, marking the spin changes between up and down two states, see Figure 2b. Magnetization measurements show that the sample goes back to its original AFM behavior after releasing the pressure (Figure S5). Based on the experimental observations in both the magneto-transport and magnetization measurements, we conclude that an FM phase transition indeed occurs in MnBi$_6$Te$_{10}$ as pressure increases.

The magnetic ground states of the layered MnBi$_2$Te$_4$(Bi$_2$Te$_3$)$_n$ compounds are the results of both interlayer and intralayer exchange coupling. While the interlayer coupling is dictated by the super-exchange between Mn-3d orbitals in adjacent layers mediated by the p orbitals of Bi and Te atoms in between, the intralayer coupling is determined by the competition between the super-exchange and direct exchange, as shown in Figure 3a. The former is between two adjacent Mn atoms mediated by a Te atom with a calculated Mn-Te-Mn bond angle of about 95°. Thus, the super-exchange coupling prefers FM state according to the Goodenough-Kanamori-Anderson (GKA) rule[40, 41]. The latter prefers AFM coupling because the direct FM hopping between Mn atoms is forbidden for Mn$^{2+}$ high-spin state ($d^5$). Because of the large Mn-Mn distance (about 4.4 Å), the direct exchange coupling is much smaller than the super-exchange. Therefore, the intralayer FM state is dominant in MnBi$_2$Te$_4$(Bi$_2$Te$_3$)$_n$.

To understand the different magnetic behaviors of MnBi$_4$Te$_7$ and MnBi$_6$Te$_{10}$ under pressure, we employ Heisenberg model to study the interlayer and intralayer exchange coupling parameters in these materials (details can be found in the supporting information). Our simulation uses three different magnetic configurations: G-type AFM, A-type AFM, and FM (Figure 3b) to evaluate the strength of the interlayer and intralayer couplings, which can be expressed by the total energy difference $E_{FM}$ - $E_{A\text{-}AFM}$, and $E_{G\text{-}AFM}$ - $E_{A\text{-}AFM}$, respectively. Our calculation shows that the hydrostatic pressure causes the lattice to shrink both along and perpendicular to the stacking direction (Figure S9). However, hydrostatic pressure monotonically increases the strength of interlayer exchange coupling but decreases the strength of intralayer coupling by about 10% (Figure 3c). The weakened intralayer coupling can be explained by the competition between direct exchange and super-exchange. In our calculation, when the pressure increases to 3 GPa, both the Mn-Te bond length and the distance between Mn atoms decrease by about 2%, while the Mn-Te-Mn bond angles are almost unchanged. Thus, the AFM-preferred direct exchange, which has a higher power of correlation to the distance (~ $d_{Mn-Mn}^{-5}$)[42, 43], will increase much more than super-exchange. Considering the decreasing intralayer FM coupling, the effects of fluctuations, including magnetic domains, thermal fluctuations, and other magnetic perturbations, become more pronounced as pressure increases. The combined effect will reduce intralayer magnetization, causing a decreased magnetic transition temperature and anomalous

Hall resistivity for both MnBi$_4$Te$_7$ and MnBi$_6$Te$_{10}$ in Figure 1.

The distinct behavior of $B_s$ and magnetic ground state for different MnBi$_2$Te$_4$(Bi$_2$Te$_3$)$_n$ under pressure can thus be understood by investigating the change of interlayer and intralayer exchange coupling for different numbers of Bi$_2$Te$_3$ spacing layers n. The intralayer magnetization decreases similarly for MnBi$_4$Te$_7$ (n = 1) and MnBi$_6$Te$_{10}$ (n = 2), as DFT calculation shows weakened intralayer FM coupling with increasing pressure (Figure 3c). The decreased total magnetic moment ($m_s$) from one magnetic septuple layer leads to a negative contribution to the interlayer exchange coupling (IEC ∝ $m_s$)[33, 44]. Such a reduced intralayer magnetization effect will not be captured by the DFT calculated change of interlayer coupling because DFT calculation considers a perfect unit cell with certain magnetic configurations at zero temperature. Considering that DFT calculated interlayer coupling is on the order of $10^{-1}$ meV for MnBi$_4$Te$_7$ and $10^{-2}$ meV for MnBi$_6$Te$_{10}$, the reduced intralayer magnetization could play a more significant role in effectively reducing the interlayer coupling in MnBi$_6$Te$_{10}$ than MnBi$_4$Te$_7$. For pressurized MnBi$_4$Te$_7$, the experimentally increased $B_s$ and strongly suppressed hysteresis loops under high pressure indicate enhanced interlayer AFM coupling. On the other hand, for pressurized MnBi$_6$Te$_{10}$, $B_s$ decreases over pressure, and a quasi-2D FM state emerges after the weak AFM kinks disappear around 1.5 GPa, consistent with our scenario of reduced and eventually decoupled magnetic layers due to reduced intralayer FM coupling. Coincidentally, the saturation field of the pressure-induced FM phase in MnBi$_6$Te$_{10}$ is also comparable to what was observed in the FM axion insulator MnBi$_8$Te$_{13}$[30, 31]. To confirm whether the lattice compression alters the topological nature of the band structure, we have performed theoretical calculations and the details are listed in the supporting information (Figure S10 and Table S1). Under 0 GPa and 2 GPa, both MnBi$_4$Te$_7$ and MnBi$_6$Te$_{10}$ exhibit robust topological nontrivial phases. Therefore an FM topological state is expected in MnBi$_6$Te$_{10}$ when pressure tunes its magnetic phase to FM.

To further validate our understanding of the competing intralayer exchange coupling, we investigate the magnetic properties of Sb doped MnBi$_6$Te$_{10}$, where the in-plane lattice constants are effectively reduced while the interlayer distance (c-axis lattice parameter) stays almost unchanged[45-49]. Compared with the pressurized case, since there is little lattice change along c-axis to enhance the interlayer coupling, the Sb-doped samples would enter the FM phase at a smaller intralayer lattice compression. We target 20% Sb doped MnBi$_6$Te$_{10}$ sample, as the in-plane lattice constant change is near but slightly lower than that of the pressurized MnBi$_6$Te$_{10}$ near the phase transition from AFM to quasi-2D-FM state. The Sb doped MnBi$_6$Te$_{10}$ crystals are prepared by replacing the Bi component with a molar ratio 1/4 of Sb to Bi in the original recipe. The actual composition is found to be Mn$_{0.77}$Sb$_{1.46}$Bi$_{4.54}$Te$_{9.54}$ by EDX, determined from the average of 18 randomly selected spots on three samples. The actual Sb concentration is about 1.2 times of nominal ratio, similar to Sb-doped MnBi$_4$Te$_7$[48]. Powder and single-crystal XRD results (Figure S11) show a 0.65% contraction of the in-plane lattice constant, while the out-of-plane lattice component stays around 101.985(8)Å (with a negligible increase of 0.07%). Interestingly, Figure 4a shows a large bifurcation of ZFC and FC curves at a magnetic ordering temperature of 11.3 K, which is similar to the FM behavior of MnBi$_8$Te$_{13}$[30, 31] and pressurized MnBi$_6$Te$_{10}$ at 1.98 GPa (Figure 2a), but in contrast with the sharp cusp AFM feature of the parent compound[27-29]. Moreover, the magnetic susceptibility of Sb doped MnBi$_6$Te$_{10}$ is one order larger than that of the parent. The magnetization curves with H//ab and H//c at different temperatures in Figure 4b also reveal a similar FM hysteresis loop as observed in the anomalous Hall effect of the parent at 1.98 GPa.

The observation of such an FM state in Sb doped MnBi$_6$Te$_{10}$ strongly supports our scenario that reduced intralayer FM coupling due to in-plane lattice compression can trigger the AFM to quasi-2D-FM phase transition in MnBi$_6$Te$_{10}$, similar to the pressurized case. It is worth noting that Sb doping may also promote Mn$^{Sb}$ antisite defects since the electronegativity and ionic size of Sb are closer to Mn than that of Bi. The exchange coupling between the original Mn and Mn occupied on Bi/Sb site within MnBi$_2$Te$_4$ layer is reported to be ferrimagnetic coupling[46, 48, 50], which can also affect the magnetism. Considering the migration of Mn and other defects, the competition between intralayer and interlayer coupling can be more complicated, leading to complex magnetic phase transitions observed in MnSb$_2$Te$_4$, and heavily Sb doped MnBi$_4$Te$_7$[48-50]. However, in our case, the Sb component is relatively small. The density of Mn$^{Sb/Bi}$ antisite defects is comparable to the parent (1%~2.1%) on the second atomic layer of the three terminations (Table S2), revealed by the scanning tunneling microscope images (Figure S13). We believe that the ferrimagnetic coupling induced by Sb doping does not play a dominant role in changing the magnetic properties. Unlike the coexistence of AFM and FM phases in Sb-doped MnBi$_4$Te$_7$ for a much higher doping level $x \sim 0.48$[48], both the *M-T* and *M-H* curves of our 20% Sb-doped MnBi$_6$Te$_{10}$ show no signature of AFM phase. In addition, comparing with the reported ferrimagnetic behaviors in MnSb$_2$Te$_4$[46, 47, 50] and the heavily Sb doped MnBi$_4$Te$_7$ (83%)[48], our 20% Sb-doped MnBi$_6$Te$_{10}$ shows a linear $1/\chi \sim T$ curve at paramagnetic state. The higher FC magnetic susceptibility (58 emu mol$^{-1}$ Oe$^{-1}$) and saturation moment at 2 K (2.84 $\mu_B$/Mn) are both close to those of the FM MnBi$_8$Te$_{13}$ (3.1 $\mu_B$/Mn)[30, 31]. Therefore, the successful realization of quasi-2D-FM state in Sb doped MnBi$_6$Te$_{10}$ provides additional evidence that the intralayer coupling plays an important role in the magnetism of weakly coupled magnetic topological insulator heterostructures.

In summary, we have systematically studied the magnetic and magneto-transport properties of MnBi$_2$Te$_4$(Bi$_2$Te$_3$)$_n$ for n = 1 and 2 under hydrostatic pressure up to 3.5 GPa. For n = 1, the saturation field increases, and the Néel temperature decreases as increasing pressure due to an enhanced interlayer AFM coupling competing combined with a weakened intralayer FM coupling. For n = 2, the interlayer AFM coupling is weak enough, so that as pressure increases, the decreased intralayer FM exchange coupling can effectively reduce the interlayer exchange coupling, resulting in a magnetic phase transition from A-type AFM to a quasi-2D FM states at around 1.5 GPa. Our results show that the intralayer exchange coupling plays a significant role in determining the magnetic properties of weakly coupled MnBi$_2$Te$_4$(Bi$_2$Te$_3$)$_n$. The intralayer exchange coupling can be delicately tuned by lattice engineering, such as pressure and chemical substitution. We show that an intrinsic FM state can be realized in both pressurized MnBi$_6$Te$_{10}$ and Sb-doped MnBi$_6$Te$_{10}$. Our results shed light on the intriguing magnetism in the family of MnBi$_2$Te$_4$(Bi$_2$Te$_3$)$_n$ and open up opportunities for the realization of various topological phases determined by their magnetic phases.

**Methods**

The single crystals were grown by flux method and confirmed by x-ray diffraction. Subsequent screening using magnetic and magneto-transport measurements were performed at physical property measurement system (PPMS). Hydrostatic pressure was applied by a self-clamped BeCu-NiCrAl double-wall piston-cylinder cell with a maximum pressure of 3.5 GPa, using Daphne 7373 oil as a pressure transmitting medium. The real Sb doping ratio was measured by energy-dispersive X-ray spectroscopy (EDX). The scanning tunneling microscope (STM) measurements were performed on *in-situ* cleaved surfaces of Mn(Bi$_{1-x}$Sb$_x$)$_6$Te$_{10}$ using commercial STM (Unisoku USM 1300) operating at 77K.

## ASSOCIATED CONTENT

**Supporting Information**

The Supporting Information is available free of charge on the ACS Publications website at DOI: **.****/acs.nanolett.*******.

Single crystal growth method and structure characterization of $MnBi_4Te_7$ and $MnBi_6Te_{10}$; Magnetic and transport properties of the crystals at ambient pressure; Measurement details of the magneto-transport properties under high pressure; Magnetic susceptibility and magnetization of $MnBi_6Te_{10}$ under high pressure measured by pre-calibrated Hall sensor; DFT calculation method; Calculated lattice compression and topological properties under pressure for both $MnBi_4Te_7$ and $MnBi_6Te_{10}$; Crystal structure and magnetic properties of 20% Sb-doped $MnBi_6Te_{10}$; Defect statistics of $Mn(Bi_{1-x}Sb_x)_6Te_{10}$ (x = 0 and 0.2) measured by scanning tunneling microscope.

## AUTHOR INFORMATION

**Corresponding Authors**

  **Yue Zhao** − *Shenzhen Institute for Quantum Science and Engineering and Department of Physics, Southern University of Science and Technology, Shenzhen 518055, China;* orcid.org/0000-0002-9174-0519; E-mail: zhaoy@sustech.edu.cn.

  **Qihang Liu** − *Shenzhen Institute for Quantum Science and Engineering and Department of Physics, Southern University of Science and Technology, Shenzhen 518055, China; Guangdong Provincial Key Laboratory for Computational Science and Material Design, Southern University of Science and Technology, Shenzhen 518055, China; Shenzhen Key Laboratory of for Advanced Quantum Functional Materials and Devices, Southern University of Science and Technology, Shenzhen 518055, China;* orcid.org/0000-0001-9843-2482; E-mail: liuqh@sustech.edu.cn.

**Authors**

  **Jifeng Shao** − *Shenzhen Institute for Quantum Science and Engineering and Department of Physics, Southern University of Science and Technology, Shenzhen 518055, China;* orcid.org/0000-0001-6460-3423

  **Yuntian Liu** − *Shenzhen Institute for Quantum Science and Engineering and Department of Physics, Southern University of Science and Technology, Shenzhen 518055, China;* orcid.org/0000-0002-7528-3995

  **Meng Zeng** − *Shenzhen Institute for Quantum Science and Engineering and Department of Physics, Southern University of Science and Technology, Shenzhen 518055, China;* orcid.org/0000-0002-5432-492X

  **Jingyuan Li** − *Shenzhen Institute for Quantum Science and Engineering and Department of*

*Physics, Southern University of Science and Technology, Shenzhen 518055, China;* orcid.org/0000-0003-3226-1822

 **Xuefeng Wu** − *Shenzhen Institute for Quantum Science and Engineering and Department of Physics, Southern University of Science and Technology, Shenzhen 518055, China*

 **Xiao-Ming Ma** − *Shenzhen Institute for Quantum Science and Engineering and Department of Physics, Southern University of Science and Technology, Shenzhen 518055, China;* orcid.org/0000-0003-3403-4228

 **Feng Jin** − *Hefei National Laboratory for Physical Sciences at Microscale, University of Science and Technology of China, Hefei 230026, China;* orcid.org/0000-0002-5688-0364

 **Ruie Lu** − *Shenzhen Institute for Quantum Science and Engineering and Department of Physics, Southern University of Science and Technology, Shenzhen 518055, China;* orcid.org/0000-0002-0403-5337

 **Yichen Sun** − *Shenzhen Institute for Quantum Science and Engineering and Department of Physics, Southern University of Science and Technology, Shenzhen 518055, China*

 **Mingqiang Gu** − *Shenzhen Institute for Quantum Science and Engineering and Department of Physics, Southern University of Science and Technology, Shenzhen 518055, China;* orcid.org/0000-0002-2889-2202

 **Kedong Wang** − *Shenzhen Institute for Quantum Science and Engineering and Department of Physics, Southern University of Science and Technology, Shenzhen 518055, China;* orcid.org/0000-0003-1253-5603

 **Wenbin Wu** − *Hefei National Laboratory for Physical Sciences at Microscale, University of Science and Technology of China, Hefei 230026, China;* orcid.org/0000-0003-1133-0016

 **Liusuo Wu** − *Shenzhen Institute for Quantum Science and Engineering and Department of Physics, Southern University of Science and Technology, Shenzhen 518055, China;* orcid.org/0000-0003-0103-5267

 **Chang Liu** − *Shenzhen Institute for Quantum Science and Engineering and Department of Physics, Southern University of Science and Technology, Shenzhen 518055, China;* orcid.org/0000-0002-7738-743X

**Author Contributions**

∥J.S. and Y.L. contributed equally. Y.Z. and Q.L. are responsible for the project. M.Z., X.M. and R.L. performed the single crystal growth and XRD characterization. J.S. performed the whole magnetic and transport measurements at different pressures with J.L. and Y.S.. F.J. and W.W. performed the magnetic measurement at ambient pressure. Y.L. performed the DFT calculations with the help from M.G.. X.W. and K.W. performed the STM measurements. J.S., Y.L., C.L., Q.L. and Y.Z. analyzed the data. J.S., Y.L., Q.L. and Y.Z. wrote the paper, and all authors participated in the discussions of the results.

**Notes**

The authors declare no competing financial interest.

**ACKNOWLEDGMENTS**

The research was supported by the Key-Area Research and Development Program of Guangdong Province (2019B010931001), National Natural Science Foundation of China under project Nos. 11674150, 11804402, U2032218, 11974326, 11804144, 11974157, 11804342, 12074161, National Key R&D Program of China (2019YFA0704900), Guangdong Innovative and Entrepreneurial Research Team Program (2016ZT06D348 and 2017ZT07C062), the Guangdong Provincial Key Laboratory of Computational Science and Material Design (Grant No. 2019B030301001), the Science, Technology, and Innovation Commission of Shenzhen Municipality (JCYJ20160613160524999, ZDSYS20190902092905285, G02206304 and G02206404 ), the Shenzhen Key Laboratory (Grant No. ZDSYS20170303165926217), the Highlight Project (No. PHYS-HL-2020-1) of the College of Science of SUSTech, and Hefei Science Center of Chinese Academy of Sciences (Grant 2018ZYFX002). First-principles calculations were also supported by Center for Computational Science and Engineering at SUSTech. We thank Junhao Lin for valuable discussions.

Carrier Engineering in intrinsic magnetic topological insulator MnBi$_4$Te$_7$. **2020,** arXiv:2009.00039v1. arXiv.org e-Print archive. https://arxiv.org/abs/2009.00039v1 (accessed on August 31, 2020).

(50) Liu, Y.; Wang, L.-L.; Zheng, Q.; Huang, Z.; Wang, X.; Chi, M.; Wu, Y.; Chakoumakos, B. C.; McGuire, M. A.; Sales, B. C.; Wu, W.; Yan, J. Site Mixing for Engineering Magnetic Topological Insulators. *Phys. Rev. X* **2021,** 11, (2), 021033.

Figure 1. Pressure tuned magnetism of MnBi$_4$Te$_7$ and MnBi$_6$Te$_{10}$. (a-b) Field dependence of the magneto-resistivity $\rho_{xx}$ (a) and the Hall resistivity $\rho_{xy}$ (b) of MnBi$_4$Te$_7$ at T = 2 K under different pressure from 0 GPa to 3.39 GPa. The curves of magneto-resistivity at zero pressure are shifted for clarity. (e-f) Field dependence of $\rho_{xx}$ and the anomalous Hall resistivity $\rho_{xy}^A$ of MnBi$_6$Te$_{10}$ at T = 2 K under different pressure. An FM phase occurs for MnBi$_6$Te$_{10}$ at 1.98 GPa. The orange curved arrows mark the change of saturation field as pressure increases. The evolution of magnetic phase transition temperature (black square) and saturation field (red dot) with pressure are summarized for MnBi$_4$Te$_7$ (c) and MnBi$_6$Te$_{10}$ (g). The magnetic ordering temperature decreases for both crystals as increasing pressure, while the saturation field responds differently. (d) and (h) are the schematic phase diagrams of MnBi$_4$Te$_7$ and MnBi$_6$Te$_{10}$, respectively. SL and QL represent the MnBi$_2$Te$_4$ septuple layer and the Bi$_2$Te$_3$ quintuple layer.

Figure 2. Magnetic susceptibility measurements of MnBi$_6$Te$_{10}$ under pressure. (a) Temperature dependence of zero-field-cooled (ZFC) and field-cooled (FC) magnetic susceptibility of MnBi$_6$Te$_{10}$ measured by Hall sensor method at different pressures under an applied magnetic field of 100 Oe along *c*-axis. (b) Isothermal magnetization of MnBi$_6$Te$_{10}$ at 2 K at various pressure with magnetic field along *c*-axis. The curves are shifted for clarity. The green and orange curved arrows mark the evolution of magnetic ordering temperature and saturation field with increasing pressure respectively.

Figure 3. (a) Schematic of the stacking of Mn and its adjacent Te layers with top view, showing the intralayer exchange coupling within a magnetic layer of MnBi$_2$Te$_4$. The intralayer coupling is determined by the competition of the direct exchange coupling between two adjacent Mn atoms (blue) and the super-exchange coupling mediated by Te atoms (red). (b) Schematics of the magnetic configurations for A-AFM, FM, and G-AFM phases. The red arrows represent the direction of the magnetic moments. (c) Top panel: the energy difference between FM and A-AFM phases for MnBi$_4$Te$_7$ and MnBi$_6$Te$_{10}$ at different pressure; bottom panel: the energy difference between G-AFM and A-AFM phases for MnBi$_4$Te$_7$ and MnBi$_6$Te$_{10}$ as pressure changes.

Figure 4. Temperature dependence of ZFC and FC magnetic susceptibility of Sb-doped MnBi$_6$Te$_{10}$ with an applied magnetic field of 100 Oe along the *c*-axis. The inset shows the temperature-dependent field cooled inverse susceptibility at H=0.4T for H//c, and the red line is the fitting result of Curie-

Weiss law. (b) Isothermal magnetization curves of the Sb doped MnBi$_6$Te$_{10}$ below 15K for H//c and H//ab (the inset). The FM loops show no kinks or plateaus associated with AFM states up to the magnetic ordering temperature of 11.3K.

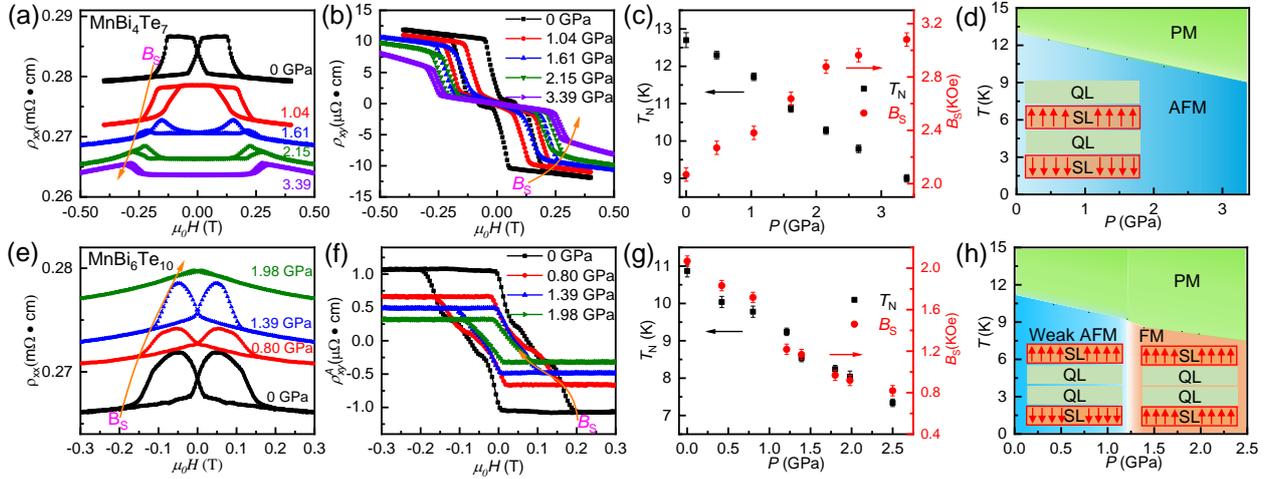

Figure 1

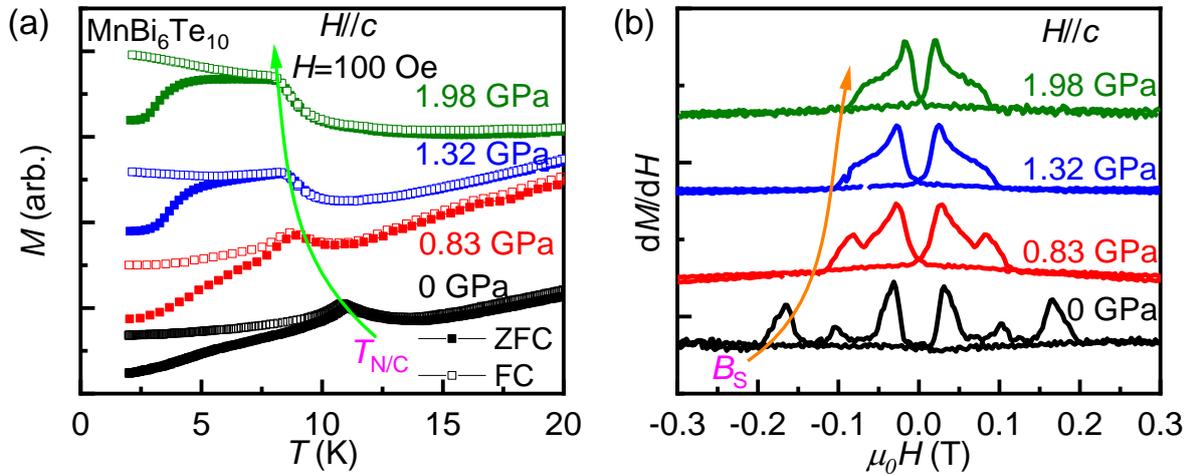

Figure 2

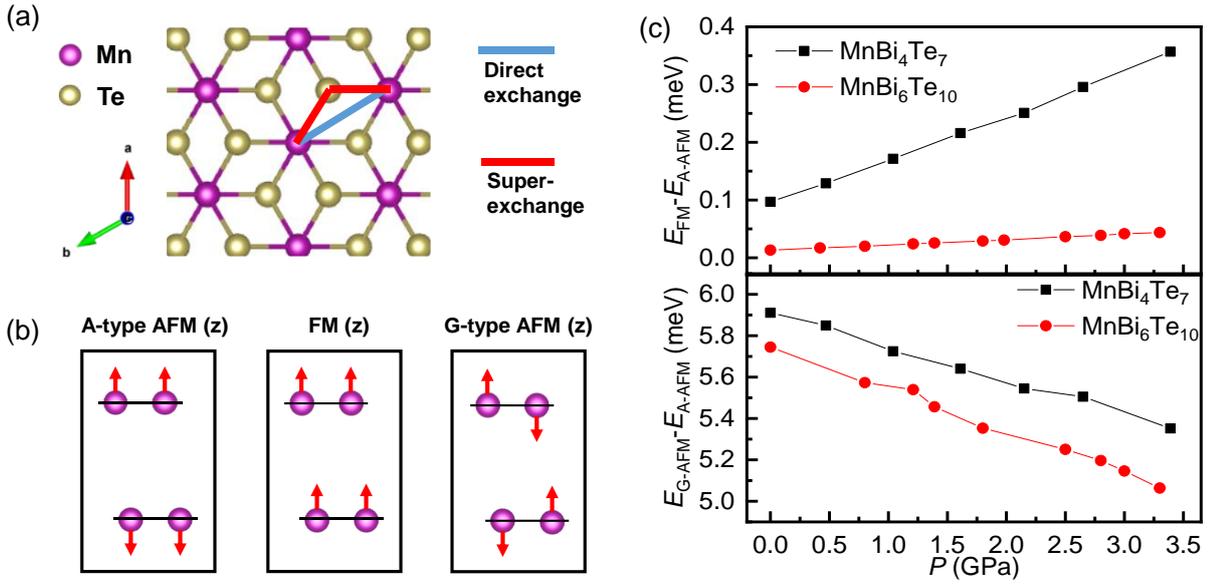

Figure 3

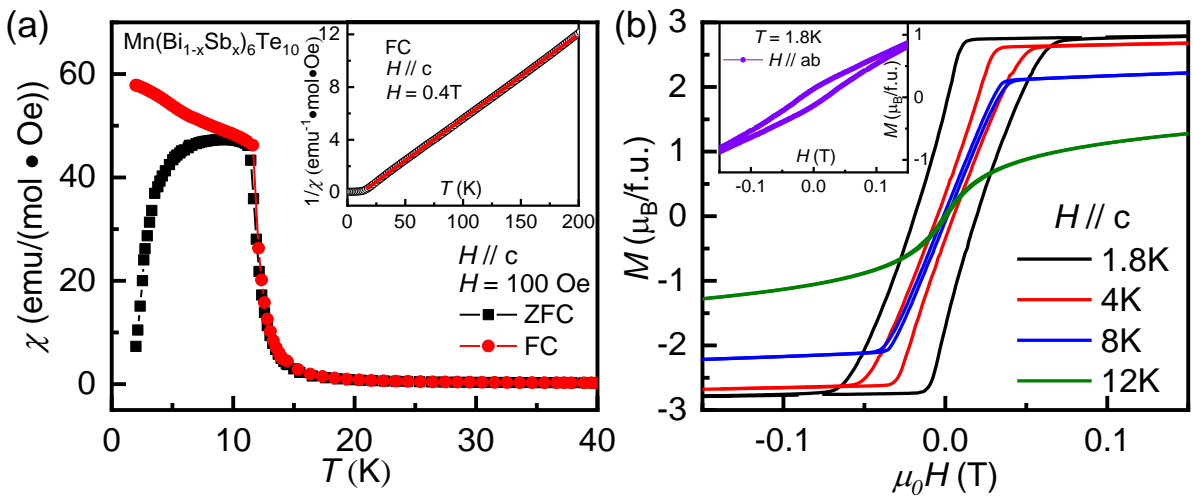

Figure 4

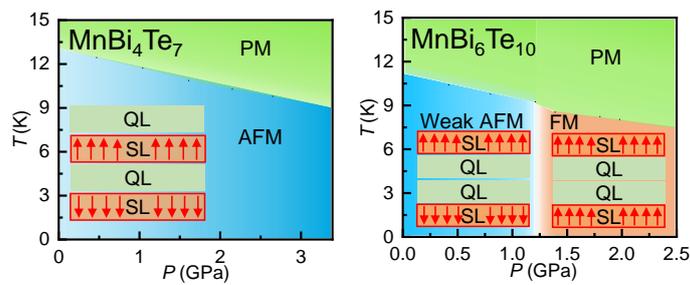

TOC graph